\documentclass[a4paper]{jpconf}

\usepackage{amsmath}
\usepackage{graphicx}
\newcommand{\hc}{\hbox {H.c.}} 

\def\lsim{\mathrel{\raise.3ex\hbox{$<$\kern-.75em\lower1ex\hbox{$\sim$}}}}
\def\gsim{\mathrel{\raise.3ex\hbox{$>$\kern-.75em\lower1ex\hbox{$\sim$}}}}
\usepackage[usenames]{color}

\allowdisplaybreaks %!!!!

\begin{document}

\setcounter{footnote}{0}
\renewcommand{\thefootnote}{\arabic{footnote}}

\title{Prospects for 2HDM charged Higgs searches}

\author{M Krawczyk$^1$, S Moretti$^2$, \underline{P Osland}$^3$, GM Pruna$^4$, R Santos$^{5,6}$}
\address{$^1$ Faculty of Physics, University of Warsaw, 
   Pasteura 5, 02-093 Warsaw, Poland}
\address{$^2$ School of Physics and Astronomy, University of Southampton,
Highfield, Southampton SO17 1BJ, United Kingdom}
\address{$^3$ Department of Physics and Technology, University of Bergen, Postboks
7803, N-5020  Bergen, Norway}
\address{$^4$ Paul Scherrer Institute, CH-5232 Villigen PSI, Switzerland}
\address{$^5$ Centro de F\'isica Te\'{o}rica e Computacional, Faculdade de Ci\^encias,
Universidade de Lisboa, Campo Grande, Edif\'icio C8 1749-016 Lisboa, Portugal, \\
$^6$ Instituto Superior de Engenharia de Lisboa - ISEL, 1959-007
Lisboa, Portugal}

\ead{Maria.Krawczyk@fuw.edu.pl, S.Moretti@soton.ac.uk, Per.Osland@uib.no, giovanni-marco.pruna@psi.ch, rasantos@fc.ul.pt}

\begin{abstract}
We discuss the prospects for charged Higgs boson searches at the LHC, within the two-Higgs-doublet models (2HDM).  The 2HDM is generally less constrained than the corresponding sector of the MSSM, but there are still severe theoretical and experimental constraints that already exclude significant regions of the naive parameter space.
Explicit searches in the $H^+\to\tau^+\nu$ and $H^+\to t\bar b$ channels are further restricting parts of the 2HDM parameter space.
\end{abstract}

%%%%%%%%%%%%%%%%%%%%%%%%%%%%%%%%%%%%%%%%%%%%%%%%%
\section{Introduction}
%%%%%%%%%%%%%%%%%%%%%%%%%%%%%%%%%%%%%%%%%%%%%%%%%
The discovery of the Higgs particle \cite{Aad:2012tfa,Chatrchyan:2012xdj} raises the obvious question of whether there are more Higgs-like particles. While the simplicity of the minimal Standard Model is very attractive, there are well-known issues that remain to be understood. So far, searches for additional neutral Higgs bosons have found nothing up to a mass of the order of 1~TeV (see, for example, Ref.~\cite{Canelli}).

The charged-Higgs sector of the 2HDM has many features in common with the MSSM, but it is a priori less constrained. 
The mass is essentially unknown, but constrained relative to other parameters of the model.
Compared to the MSSM, there is more room for a wide range of masses. In particular, the charged Higgs boson mass could a priori be rather different from the neutral Higgs masses.

In the next section we shall introduce some basic notation, and review the most important constraints. Then, in the following sections we will separately discuss three mass ranges. In section~\ref{sect:low-mass} we discuss the light-mass case, where a top quark may decay to a charged Higgs boson, $t\to H^+b$. In section~\ref{sect:intermediate-mass} we discuss the intermediate mass region, where the charged Higgs boson is too heavy to be produced via top decay, but where the Model~II is excluded by the $b\to X_s\gamma$ constraint, up to of the order of 480~GeV \cite{Misiak:2015xwa}\footnote{A recent study concludes that this limit is even higher, at 570~GeV \cite{Misiak:2017bgg}.}. In section~\ref{sect:high-mass} we consider still higher masses, where the $b\to X_s\gamma$ constraint does not apply.
For a complementary review, see Ref.~\cite{Moretti:2016qcc}.

%%%%%%%%%%%%%%%%%%%%%%%%%%%%%%%%%%%%%%%%%%%%%%%%%
\section{Model review}
\setcounter{equation}{0}
%%%%%%%%%%%%%%%%%%%%%%%%%%%%%%%%%%%%%%%%%%%%%%%%%
The 2HDM is a simple extension of the SM, described in detail in Refs.~\cite{Gunion:1989we,Branco:2011iw}.
%%%%%%%%%%%%%%%%%%%%%%%%%%%%%%%%%%%%%%%%%%%%%%%%%
\subsection{Notations}
%%%%%%%%%%%%%%%%%%%%%%%%%%%%%%%%%%%%%%%%%%%%%%%%%
We decompose the SU(2) doublets as follows:
\begin{equation} \label{Eq:Obasis}
\Phi_a=
\begin{pmatrix}
\varphi_a^+\\ (v_a+\eta_a+i\chi_a)/\sqrt{2}
\end{pmatrix}, \quad a=1,2
\end{equation}
with the vacuum expectation value (vev) $v_a$ real. The ratio between these is denoted $\tan\beta=v_2/v_1$, and plays an important role in the parametrization of the model.

The potential is written as
\begin{align}
\label{Eq:pot}
V(\Phi_1,\Phi_2) &= -\frac12\left\{m_{11}^2\Phi_1^\dagger\Phi_1
+ m_{22}^2\Phi_2^\dagger\Phi_2 + \left[m_{12}^2 \Phi_1^\dagger \Phi_2
+ \hc\right]\right\} \nonumber \\
& + \frac{\lambda_1}{2}(\Phi_1^\dagger\Phi_1)^2
+ \frac{\lambda_2}{2}(\Phi_2^\dagger\Phi_2)^2
+ \lambda_3(\Phi_1^\dagger\Phi_1)(\Phi_2^\dagger\Phi_2) 
+ \lambda_4(\Phi_1^\dagger\Phi_2)(\Phi_2^\dagger\Phi_1)\nonumber \\
&+ \frac12\left[\lambda_5(\Phi_1^\dagger\Phi_2)^2 + \hc\right]
+\left\{\left[\lambda_6(\Phi_1^\dagger\Phi_1)+\lambda_7
(\Phi_2^\dagger\Phi_2)\right](\Phi_1^\dagger\Phi_2)
+{\rm \hc}\right\}.
\end{align}
In many studies, the terms with $\lambda_6$ and $\lambda_7$ are left out, in order to control flavour-changing neutral interactions.

The spectrum consists of three neutral Higgs boson (the lightest one is here assumed to be the discovered one, at 125~GeV), and a charged pair, $H^\pm$. If CP is conserved, there are two CP-even ones ($h$ and $H$, arising from $\eta_1$ and $\eta_2$ in Eq.~(\ref{Eq:Obasis})), and an odd one ($A$, arising, together with the neutral Goldstone boson, from $\chi_1$ and $\chi_2$). If, on the other hand CP is not conserved, they will mix to mass eigenstates $H_1$, $H_2$ and $H_3$, with $H_1$ being the lightest one.

The Yukawa interactions can be written as
\begin{equation}
-\mathcal{L}_{\mathrm{Yukawa}}
=\overline{Q}_{L}\Phi_aF_a^DD_R+\overline{Q}_{L}\widetilde{\Phi}_aF_a^UU_R+\overline{L}_{L}\Phi_aF_a^LL_R+\hc, \quad a=1,2
\end{equation}
where $\widetilde{\Phi}_a=i\sigma_2\Phi^\ast$ and $Q_L$ and $L_L$ are left-handed quark and lepton doublets, whereas $D_R$, $U_R$ and $L_R$ are the corresponding right-handed fields.

Different choices for the $F_a$ define different ``Models'': Model~I is defined by $\Phi_2$ (and only $\Phi_2$) coupling to all fermions, whereas in Model~II the $d$-type quarks and charged leptons couple exclusively to $\Phi_1$, whereas $u$-type quarks couple exclusively to $\Phi_2$. This is similar to the structure of the MSSM \cite{Djouadi:2005gj}. These coefficients $F_a$ are proportional to the quark (and lepton) masses, as required by the Higgs mechanism.
The implications for the charged-Higgs coupling are that they will consist of two parts, one with a left-handed and the other with a right-handed projector, one proportional to the down-type mass, the other to the up-type mass. The different models give different combinations of projectors and mass factors. For example, for Model~II and third-generation quarks we have (with all fields incoming)
\begin{alignat}{2}  \label{Eq:Yukawa-charged-II}
&H^+ b \bar t: &\qquad
&\frac{ig}{2\sqrt2 \,m_W}\,V_{tb}
[m_b(1+\gamma_5)\tan\beta+m_t(1-\gamma_5)\cot\beta], \nonumber \\
&H^-  t\bar b: &\qquad
&\frac{ig}{2\sqrt2 \,m_W}\,V_{tb}^*
[m_b(1-\gamma_5)\tan\beta+m_t(1+\gamma_5)\cot\beta],
\end{alignat}
where $g$ is the SU(2) coupling and $V$ the CKM matrix.

There are also Models~X and Y, which are basically variants of Models~I and II, with different leptonic couplings.
%%%%%%%%%%%%%%%%%%%%%%%%%%%%%%%%%%%%%%%%%%%%%%%%%
\subsection{Important constraints}
%%%%%%%%%%%%%%%%%%%%%%%%%%%%%%%%%%%%%%%%%%%%%%%%%
A recent summary of theoretical and experimental constraints can be found in Ref.~\cite{Akeroyd:2016ymd}.
The theoretical constraints include positivity, unitarity and perturbativity, whereas the experimental ones arise from low-energy phenomena, such as various $B$-meson decays where the existence of a charged Higgs boson would contribute additional effects, as well as precision measurements at LEP constraining radiative corrections related to the electroweak bosons $W$ and $Z$.

Among the constraints coming from $B$-meson decay, the $B\to X_s\gamma$ decay, which is experimentally well measured and in the SM dominated by a one-loop diagram involving the $W^+$ will in the 2HDM Model~II also get a contribution from $H^+$ exchange. Careful analyses of this effect have led to a mass bound $M_{H^\pm}\gsim480~\text{GeV}$ \cite{Misiak:2015xwa} (recently updated to $M_{H^\pm}\gsim570~\text{GeV}$ \cite{Misiak:2017bgg}), largely independent of $\tan\beta$.

The relevant LEP precision data, derived from the vacuum polarization, or radiative corrections to the $W$ and $Z$ self-energies are usually quantified in terms of the parameters $S$, $T$ and $U$  \cite{Kennedy:1988sn,Peskin:1990zt,Altarelli:1990zd}. If the charged Higgs is heavy, the smallness of $T$ \cite{Agashe:2014kda} imposes strong constraints on the allowed splitting between $M_{H^\pm}$, $M_H$ and $M_A$ \cite{Grimus:2007if,Grimus:2008nb}.
%%%%%%%%%%%%%%%%%%%%%%%%%%%%%%%%%%%%%%%%%%%%%%%%%
\section{The low-mass region, $M_{H^\pm}\lsim m_t$}
\setcounter{equation}{0}
\label{sect:low-mass}
%%%%%%%%%%%%%%%%%%%%%%%%%%%%%%%%%%%%%%%%%%%%%%%%%
In the low-mass region, where the $H^+$ can be produced via $t$ decay, Model~II and Model~Y are excluded by the $B\to X_s\gamma$ constraint. For Model~I, all Yukawa couplings are proportional to $\cot\beta$. Thus, the sensitivity is larger at low $\tan\beta$.

The LHC experiments have already set very strong bounds on the branching ratio for $t\to H^+b$, assuming the decay $H^+\to \tau^+\nu$ \cite{Aad:2014kga,Khachatryan:2015qxa}. These bounds effectively exclude some range of low values of $\tan\beta$, as illustrated in Fig.~\ref{Fig:br-top-limit}, where the region {\it above} the dashed bars is excluded by the two experiments at the 95\% CL. 

%%%%%%%%%%%%%%%%%%%%%%%%%%%%%%%%%%%%%%%%%%%%%%%%%%%%%%%%%%%%%%%%
\begin{figure}[htb]
\refstepcounter{figure}
\label{Fig:br-top-limit}
\addtocounter{figure}{-1}
\begin{center}
\includegraphics[scale=0.60]{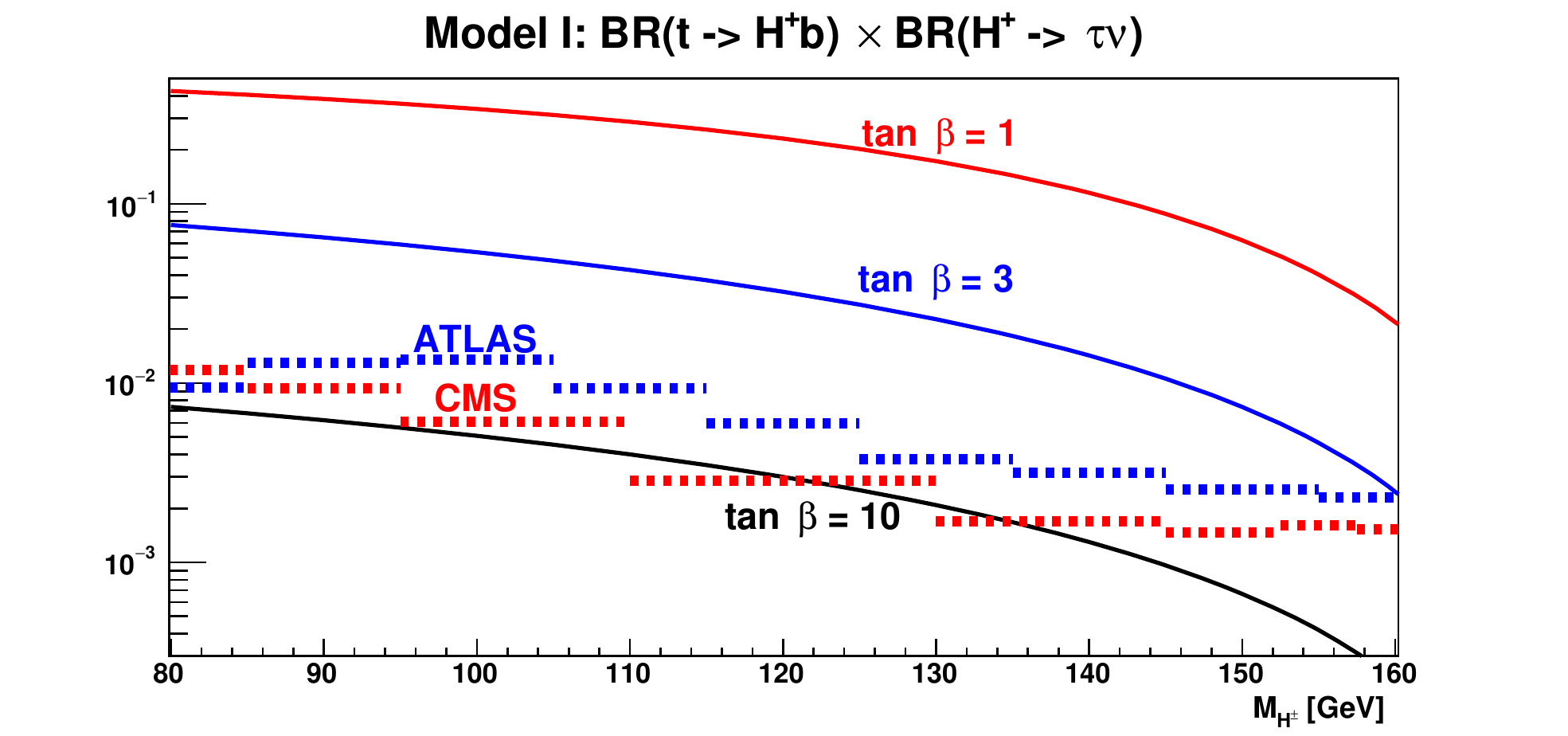}
\end{center}
\vspace*{-4mm}
\caption{Product of branching ratios, $\text{BR}(t\to H^+b)\times\text{BR}(H^+\to\tau^+\nu)$, for Model~I, and three values of $\tan\beta$, as indicated (solid). The figure is adapted from Ref.~\cite{Akeroyd:2016ymd}, showing also 8~TeV exclusion limits from ATLAS \cite{Aad:2014kga} (dashed, blue) and CMS \cite{Khachatryan:2015qxa} (dashed, red).}
\end{figure}
%%%%%%%%%%%%%%%%%%%%%%%%%%%%%%%%%%%%%%%%%%%%%%%%%%%%%%%%%%%%%%%%

It is seen that in the low-mass region, only Model~I with high values of $\tan\beta$, $\tan\beta\gsim{\cal O}(5)$ is still allowed.

Model~X is like Model~I in its couplings to quarks, and would thus have the same branching ratio for $t\to H^+b$. However, the leptonic coupling is proportional to $\tan\beta$ rather than $\cot\beta$. Thus, at low values of $\tan\beta$ the exclusion in Model~X is somewhat less severe than in Model~I.

In this mass region, assuming the $H^+\to W^+A/H$ and $H^+\to W^+h$ rates to be negligible (due to phase space and couplings, respectively), we can approximate the $H^+\to \tau^+\nu$ branching ratio as
\begin{equation}
\text{BR}(H^+\to \tau^+\nu)\simeq
\frac{\Gamma(H^+\to \tau^+\nu)}{\Gamma(H^+\to \tau^+\nu)+\Gamma(H^+\to c\bar s)}.
\end{equation}
In Model~I, with both rates proportional to $\cot^2\beta$, the branching ratio is independent of $\tan\beta$ and of the order of 0.7 \cite{Akeroyd:2016ymd}, whereas in Model~X the rate $\Gamma(H^+\to \tau^+\nu)$ is instead proportional to $\tan^2\beta$ and the branching ratio quickly approaches unity. Thus, the deterioration in exclusion reach from Model~I to Model~X is rather modest.

In the decay of $t\to H^+b$, one has also searched for $H^+\to c\bar s$ \cite{Khachatryan:2015uua}, but the Model~I branching ratio is lower by a factor $\sim 3/7$ and this channel is not compatible with the $\tau\nu$ channel.

%%%%%%%%%%%%%%%%%%%%%%%%%%%%%%%%%%%%%%%%%%%%%%%%%
\section{The intermediate-mass region, $m_t\lsim M_{H^\pm}\lsim 480~\text{GeV}$}
\setcounter{equation}{0}
\label{sect:intermediate-mass}
%%%%%%%%%%%%%%%%%%%%%%%%%%%%%%%%%%%%%%%%%%%%%%%%%
Also in the intermediate-mass region, Model~II and Model~Y are excluded by the $B\to X_s\gamma$ constraint. But in contrast to the low-mass region, the charged Higgs boson can not be produced via top decay. A variety of production mechanisms are involved. In the 5-flavour scheme\footnote{For a complete discussion on the flavour scheme choice in inclusive charged Higgs production associated with fermions see IV.3.2 of \cite{deFlorian:2016spz} and references therein.}, where $b$ quarks are considered parts of the proton and have non-zero distribution functions, the relevant partonic processes are
\begin{equation} \label{Eq:production}
g\bar b\to H^+ \bar t, \quad 
gg\to H_j\to H^+W^-.
\end{equation}
Two comments are here in order:
\begin{itemize}
\item
As indicated, the production can proceed via a neutral Higgs boson in the $s$-channel.
We recall the relevant couplings (all fields incoming)
\begin{alignat}{2}
&H^\mp W^\pm h: &\quad
&\frac{\mp ig}{2}
\cos(\beta -\alpha)
(p_\mu-p_\mu^\mp), \nonumber \\
&H^\mp W^\pm H: &\quad
&\frac{\pm ig}{2}
\sin(\beta -\alpha)
(p_\mu-p_\mu^\mp),  \nonumber \\
&H^\mp W^\pm A: &\quad
&\frac{g}{2}
(p_\mu-p_\mu^\mp),
\label{Eq:CPC-gauge-couplings}
\end{alignat}
where $p_\mu$ and $p_\mu^\mp$ refer to the momenta of the neutral and charged scalars.
Experimental data \cite{Canelli} favour the alignment limit, where
\begin{equation}
\cos(\beta-\alpha)\to0.
\end{equation}
Thus, in this limit only the heavier Higgs bosons ($H$ and $A$) can play a role here. Allowing for CP non-conservation (considering $H_1$ rather than $h$) does little to modify this conclusion.
\item
In Model~I (and Model~X) all couplings of the $H^+$ to quarks are proportional to $\cot\beta$.
Thus, the production cross section decreases with increasing values of $\tan\beta$.

This is different from the MSSM, which has a Model~II structure, and where the cross section has a minimum for intermediate values of $\tan\beta$, of the order of $\sqrt{m_t/m_b}$.
\end{itemize}

As suggested by Eq.~(\ref{Eq:production}), the kinematics may allow for $s$-channel resonant effects in the production \cite{Asakawa:2005nx}. However, the smallness of $T$ does not allow for a large mass splitting between $H^\pm$, $H$ and $A$. This restricts the possible resonant enhancement.

Likewise, there are various possible decay modes:
\begin{equation} \label{Eq:decaymodes}
H^+\to \tau^+ \nu, \quad 
H^+\to W^+ h, \quad
(\text{or }H^+\to W^+ H_1), \quad
H^+\to t \bar b
\end{equation}
As mentioned above, the $H^+ \to W^+ h$ vanishes in the alignment limit. For the more general case, allowing for CP violation, the relevant coupling is given by
\begin{equation} \label{Eq:H_chWH_1}
H^\mp W^\pm H_1: \qquad
\frac{g}{2}
[\pm i\cos\alpha_2\sin(\beta-\alpha_1)+ \sin\alpha_2]
(p_\mu-p_\mu^\mp).
\end{equation}
Here, $\alpha_1=\alpha+\pi/2$ and $\alpha_2$ are two of the rotation angles describing mixing of the neutral Higgs fields \cite{ElKaffas:2006gdt}. A non-zero value of $\alpha_2$ would indicate CP violation.

The decay $H^+\to W^+h$ (or $H^+\to W^+H_1$) requires some deviation from the alignment limit, whereas $H^+\to W^+A$ and $H^+\to W^+A$ are limited by phase space. Thus, in practice, the channels $H^+\to \tau^+\nu$ and $H^+\to t\bar b$ are the most relevant ones, with the latter dominating by a $\tan\beta$-independent factor of the order of $(m_t/m_\tau)^2$.
%%%%%%%%%%%%%%%%%%%%%%%%%%%%%%%%%%%%%%%%%%%%%%%%%
\subsection{The $\tau\nu$ channel}
%%%%%%%%%%%%%%%%%%%%%%%%%%%%%%%%%%%%%%%%%%%%%%%%%
The tau channel benefits from being rather ``clean'', and interesting limits have been obtained. Considering as an example, the case $M_{H^\pm}=300~\text{GeV}$, we note that both ATLAS and CMS obtain a 95\%~CL bound $\sigma\cdot\text{BR}(H^+\to\tau^+\nu)\leq (0.16-0.17)~\text{pb}$ at 8~TeV \cite{Aad:2014kga,Khachatryan:2015qxa}. A recent ATLAS result at 13~TeV is 0.41~pb.\footnote{The charged-Higgs production cross section roughly scales with the glue-glue luminosity, which in the mass range 200-400~GeV increases by factors of 4--5 from 8 to 13~TeV.} This is to be compared with the model expectation for the cross section and the branching ratio. Approximate 14~TeV cross sections are quoted in Table~\ref{Table:modelI-sigma} for a few $\tan\beta$ values. Accurate values depend on the masses of the heavier neutral states. Barring any significant decay to $W^+h$, the relevant branching ratio is given by $(m_\tau/m_t)^2\sim10^{-4}$. It is seen that the theoretical expectation is far below the current experimental limit.
%%%%%%%%%%%%%%%%%%%%%%%%%%%%%%%%%%%%%%%%%%%%%
\begin{table}[htb]
\caption{Experimental limits on $\sigma\cdot\text{BR}(H^+\to\tau^+\nu)$ [pb] at 8 and 13~TeV together with Model~I cross sections \cite{Akeroyd:2016ymd} at 14~TeV, multiplied by a branching ratio taken to be $10^{-4}$. Three intermediate values of $M_{H^\pm}$  are considered, and for the theory cross sections we take $(M_2,M_3)=(500,600)~\text{GeV}$.\\ }
\label{Table:modelI-sigma}
\begin{center}
\begin{tabular}{|cccc|}
\hline
\hline
  & 200~GeV & 300~GeV & 400~GeV \\
\hline
\hline
ATLAS \cite{Aad:2014kga}, 8~TeV & 0.47 & 0.17 & 0.061 \\
\hline
CMS \cite{Khachatryan:2015qxa}, 8~TeV & 0.36 & 0.16 & 0.054 \\
\hline
ATLAS \cite{Aaboud:2016dig}, 13~TeV & 1.83 & 0.41 & 0.20 \\
\hline
$\tan\beta=1$, 14~TeV & $1.18\times10^{-3}$  & $0.646\times10^{-3}$  & $0.287\times10^{-3}$ \\
\hline
$\tan\beta=3$, 14~TeV & $1.41\times10^{-4}$ & $0.882\times10^{-4}$ & $0.514\times10^{-4}$ \\
\hline
$\tan\beta=30$, 14~TeV & $1.42\times10^{-6}$ & $0.91\times10^{-6}$ & $0.72\times10^{-6}$ \\
\hline
\hline
\end{tabular}
\end{center}
\end{table}
%%%%%%%%%%%%%%%%%%%%%%%%%%%%%%%%%%%%%%%%%%%%%

Furthermore, as long as the relevant search region is around $\tan\beta=1$, there is no significant difference between Model~I and Model~X, so Model~X is no more and no less affected by the experimental bounds on this channel than Model~I.

%%%%%%%%%%%%%%%%%%%%%%%%%%%%%%%%%%%%%%%%%%%%%%%%%%%%%%%%%%%%%%%%
\begin{figure}[htb]
\refstepcounter{figure}
\label{Fig:cms-intermediate-mass-limit}
\addtocounter{figure}{-1}
\begin{center}
\includegraphics[scale=0.4]{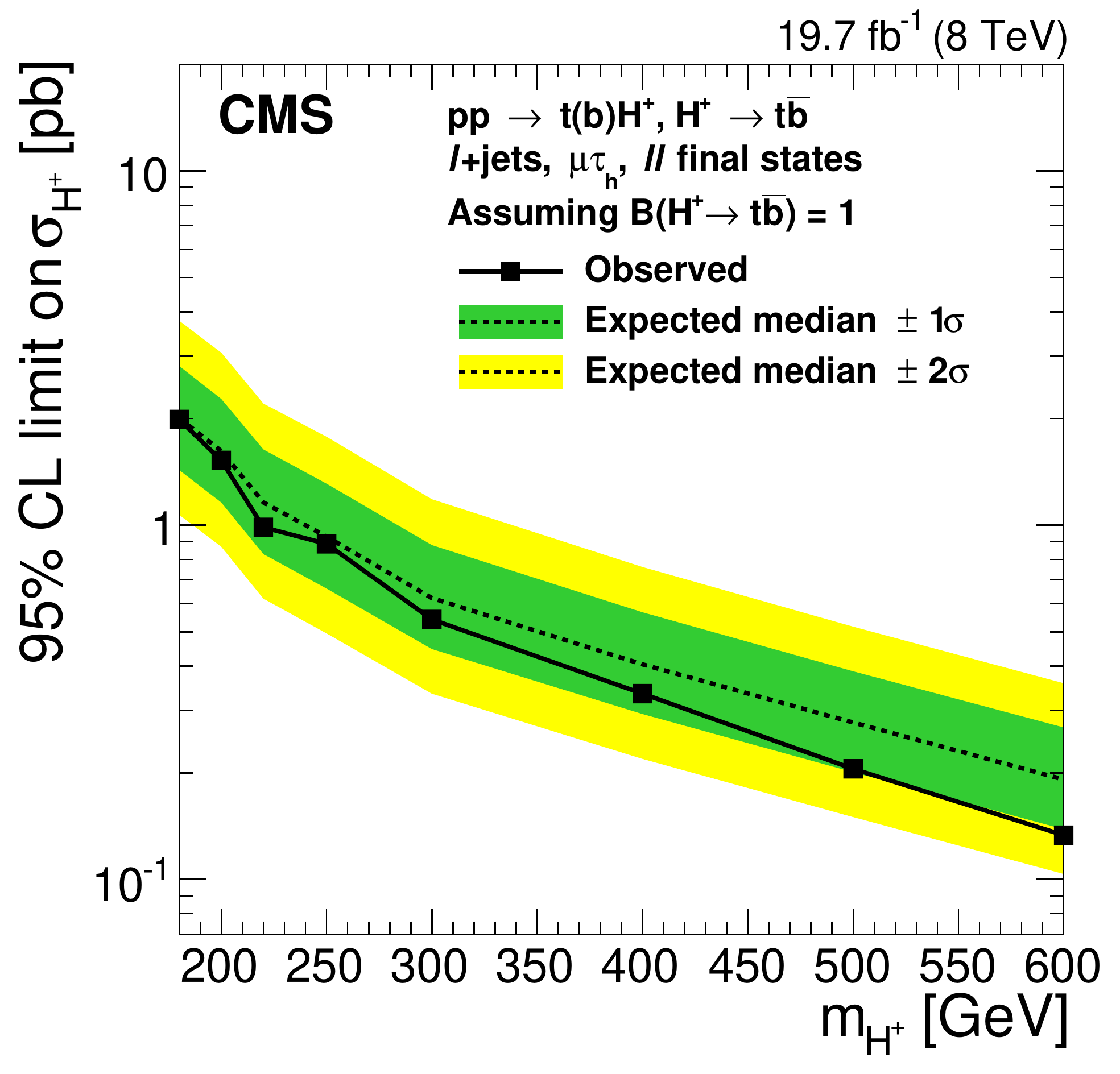}
\end{center}
\vspace*{-4mm}
\caption{CMS upper limit on $\sigma(pp\to t(b)H^+)$ for the combination of the $\mu\tau_h$, $\ell+\text{jets}$ and $\ell\ell^\prime$ final states assuming $\text{BR}(H^+\to t\bar b)=100\%$. [Reprinted with kind permission from JHEP and the authors, Fig.~10 of \cite{Khachatryan:2015qxa}.]}
\end{figure}
%%%%%%%%%%%%%%%%%%%%%%%%%%%%%%%%%%%%%%%%%%%%%%%%%%%%%%%%%%%%%%%%
%%%%%%%%%%%%%%%%%%%%%%%%%%%%%%%%%%%%%%%%%%%%%%%%%
\subsection{The $t\bar b$ channel}
%%%%%%%%%%%%%%%%%%%%%%%%%%%%%%%%%%%%%%%%%%%%%%%%%
While the $H^+\to t\bar b$ channel suffers from high QCD background rates, sophisticated analyses have started to constrain these models. 8~TeV exclusion limits from CMS are shown in Fig~\ref{Fig:cms-intermediate-mass-limit} and reproduced in Table~\ref{Table:modelI-sigma-tb}, together with preliminary ATLAS results at 13~TeV and theoretical expectations for three values of $\tan\beta$. 
For this channel, assuming $\text{BR}(H^+\to t\bar b)=100\%$, we find that low values of $\tan\beta$ are excluded. On the other hand, $\sigma(\tan\beta=30)\sim10^{-2}~\text{pb}$, which is well below the exclusion limit. 

%%%%%%%%%%%%%%%%%%%%%%%%%%%%%%%%%%%%%%%%%%%%%
\begin{table}[htb]
\caption{Experimental limits on $\sigma\cdot\text{BR}(H^+\to t\bar b)$ [pb] together with Model~I cross sections \cite{Akeroyd:2016ymd} multiplied by a branching ratio taken to be 1. Three intermediate values of $M_{H^\pm}$  are considered, and for the theory cross sections we take $(M_2,M_3)=(500,600)~\text{GeV}$.\\ }
\label{Table:modelI-sigma-tb}
\begin{center}
\begin{tabular}{|cccc|}
\hline
\hline
  & 200~GeV & 300~GeV & 400~GeV \\
\hline
\hline
CMS \cite{Khachatryan:2015qxa}, 8~TeV & 1.53 & 0.53 & 0.33 \\
\hline
ATLAS \cite{ATLAS:2016qiq}, 13~TeV &  & 1.06 & 1.17 \\
\hline
$\tan\beta=1$, 14~TeV & $11.8$  & $6.46$  & $2.87$ \\
\hline
$\tan\beta=3$, 14~TeV & $1.41$ & $0.882$ & $0.514$ \\
\hline
$\tan\beta=30$, 14~TeV & $1.42\times10^{-2}$ & $0.91\times10^{-2}$ & $0.72\times10^{-2}$ \\
\hline
\hline
\end{tabular}
\end{center}
\end{table}
%%%%%%%%%%%%%%%%%%%%%%%%%%%%%%%%%%%%%%%%%%%%%

We conclude that in this intermediate mass region the search in the $H^+\to t\bar b$ is starting to constrain the model.
Since a hadronic final state is considered in this search, there is no difference between Model~I and Model~X.

%%%%%%%%%%%%%%%%%%%%%%%%%%%%%%%%%%%%%%%%%%%%%%%%%
\section{The high-mass region, $480~\text{GeV}\lsim M_{H^\pm}$}
\setcounter{equation}{0}
\label{sect:high-mass}
%%%%%%%%%%%%%%%%%%%%%%%%%%%%%%%%%%%%%%%%%%%%%%%%%
The high-mass region is distinguished by all Yukawa models being allowed, and also all decay modes listed in Eq.~(\ref{Eq:decaymodes}).

In distinction from the low- and intermediate-mass regions discussed above, we now need to consider also the Model~II cross sections. Cross sections are quoted for both models in Table~\ref{Table:modelI-sigma-hi} \cite{Akeroyd:2016ymd}. 
Note that the cross sections for the four models are pairwise the same,
\begin{equation}
\sigma(\text{Model~X})=\sigma(\text{Model~I}), \quad \text{and} \quad
\sigma(\text{Model~Y})=\sigma(\text{Model~II}),
\end{equation}
but the decay rates would be different if lepton modes were considered.
Because of the dependence on the $H$ and $A$ (or $H_2,H_3$) masses, there is some uncertainty associated with the cross sections.
The main features are:
\begin{itemize}
\item
The cross sections fall off at high masses, mainly due to falling pdf's and phase space reduction, and
\item
The Model~I cross section falls with $\tan\beta$ approximately as $1/\tan^2\beta$, whereas the Model~II cross section has a minimum around $\tan\beta={\cal O}(\sqrt{m_t/m_b})$ due to the competition among the two terms in the coupling (\ref{Eq:Yukawa-charged-II}).
\end{itemize}

%%%%%%%%%%%%%%%%%%%%%%%%%%%%%%%%%%%%%%%%%%%%%
\begin{table}[htb]
\caption{Theoretical cross sections \cite{Akeroyd:2016ymd} for $\sqrt{s}=14~\text{TeV}$ and a range of mass values $M_{H^\pm}=500-1000~\text{GeV}$, for $(M_2,M_3)=(500,600)~\text{GeV}$, quoted in the format ``Model~I/Model~II''  [pb]. Overall factors $10^{-1}$ and $10^{-2}$ are extracted where indicated. \\ }
\label{Table:modelI-sigma-hi}
\begin{center}
\begin{tabular}{|ccccccc|}
\hline
\hline
  & 500~GeV & 600~GeV & 700~GeV & 800~GeV & 900~GeV & 1000~GeV\\
\hline
\hline
$\tan\beta=1$ & $1.10/1.27$  & $0.54/0.64$  & $0.30/0.37$ & $0.18/0.22$  & $0.11/0.14$  & $0.072/0.090$ \\
\hline
$\tan\beta=3$ $[10^{-1}]$ & $1.61/1.81$ & $0.60/0.72$ & $0.34/0.41$ & $0.20/0.25$  & $0.13/0.16$  & $0.080/0.10$ \\
\hline
$\tan\beta=30$  $[10^{-2}]$ & $0.29/24.24$ & $0.059/12.95$ & $0.033/7.31$ & $0.020/4.30$  & $0.013/2.61$  & $0.0080/1.63$ \\
\hline
\hline
\end{tabular}
\end{center}
\end{table}
%%%%%%%%%%%%%%%%%%%%%%%%%%%%%%%%%%%%%%%%%%%%%

%%%%%%%%%%%%%%%%%%%%%%%%%%%%%%%%%%%%%%%%%%%%%%%%%
\subsection{The $\tau\nu$ channel}
%%%%%%%%%%%%%%%%%%%%%%%%%%%%%%%%%%%%%%%%%%%%%%%%%
Search results from the $\tau\nu$ channel are given in Table~\ref{Table:modelI-sigma-hi-tau}. For comparison, we quote the theoretical expectation for $\tan\beta=1$, for both Model~I and Model~II. It is seen that the expectation, at $\tan\beta=1$ is lower by factors ranging from 200 to 600. For higher values of $\tan\beta$, the discrepancy would be even larger. Thus, searches in the $\tau\nu$ channel are not yet constraining the 2HDM in this mass range.

%%%%%%%%%%%%%%%%%%%%%%%%%%%%%%%%%%%%%%%%%%%%%
\begin{table}[htb]
\caption{Experimental limits on $\sigma\cdot\text{BR}(H^+\to \tau^+\nu)$ [pb] together with the theoretical expectation for $\tan\beta=1$, in the format ``Model~I/Model~II'', approximating $\text{BR}(H^+\to \tau\nu)\simeq(m_\tau/m_t)^2\simeq10^{-4}$. Overall factors $10^{-2}$ and $10^{-4}$ are extracted where indicated. \\ }
\label{Table:modelI-sigma-hi-tau}
\begin{center}
\begin{tabular}{|ccccccc|}
\hline
\hline
  & 500~GeV & 600~GeV & 700~GeV & 800~GeV & 900~GeV & 1000~GeV\\
\hline
\hline
ATLAS \cite{Aad:2014kga} $[10^{-2}]$, 8~TeV & $2.46$  & $1.11$ & & & & $0.427$ \\
\hline
ATLAS \cite{Aaboud:2016dig} $[10^{-2}]$, 13~TeV & $9.25$  & $5.15$ & $3.35$  & $2.36$ & $2.10$ & $1.80$ \\
\hline
$\tan\beta=1$ $[10^{-4}]$, 14~TeV & $1.10/1.27$  & $0.54/0.64$ & $0.30/0.37$  & $0.18/0.22$  & $0.11/0.14$  & $0.07/0.09$ \\
\hline
\hline
\end{tabular}
\end{center}
\end{table}
%%%%%%%%%%%%%%%%%%%%%%%%%%%%%%%%%%%%%%%%%%%%%

%%%%%%%%%%%%%%%%%%%%%%%%%%%%%%%%%%%%%%%%%%%%%%%%%
\subsection{The $t\bar b$ channel}
%%%%%%%%%%%%%%%%%%%%%%%%%%%%%%%%%%%%%%%%%%%%%%%%%
While the $H^+\to t\bar b$ channel suffers from large QCD backgrounds, the constraints obtained are more interesting. Experimental exclusion limits are given in Table~\ref{Table:modelI-sigma-hi-tb}, and can be directly compared with the theoretical expectations quoted in Table~\ref{Table:modelI-sigma-hi}, approximating $\text{BR}(H^+\to t\bar b)=1$.
It is seen that in the range $M_{H^\pm}=500-600~\text{GeV}$ low values of $\tan\beta$, namely $\tan\beta={\cal O}(1)$ are on the verge of being excluded for both Model~I and Model~II. Since no leptonic couplings are involved, this conclusion applies equally to Models~X and Y.

%%%%%%%%%%%%%%%%%%%%%%%%%%%%%%%%%%%%%%%%%%%%%
\begin{table}[htb]
\caption{Experimental limits on $\sigma\cdot\text{BR}(H^+\to t\bar b)$ [pb]. \\ }
\label{Table:modelI-sigma-hi-tb}
\begin{center}
\begin{tabular}{|ccccccc|}
\hline
\hline
  & 500~GeV & 600~GeV & 700~GeV & 800~GeV & 900~GeV & 1000~GeV\\
\hline
\hline
ATLAS \cite{Aad:2015typ}, 8~TeV & 0.68 & 0.24 &  & & & \\
\hline
ATLAS \cite{Aad:2015typ}, 8~TeV & 2.89 & 1.42 & 0.70 & 0.37 & 0.24 & 0.15 \\
\hline
CMS \cite{Khachatryan:2015qxa}, 8~TeV & 0.20 & 0.13 &  & & & \\
\hline
ATLAS \cite{ATLAS:2016qiq}, 13~TeV & 1.32 & 1.01 & 0.53 & 0.34 & 0.31 & 0.18 \\
\hline
\hline
\end{tabular}
\end{center}
\end{table}
%%%%%%%%%%%%%%%%%%%%%%%%%%%%%%%%%%%%%%%%%%%%%

%%%%%%%%%%%%%%%%%%%%%%%%%%%%%%%%%%%%%%%%%%%%%%%%%
\section{Summary}
\setcounter{equation}{0}
%%%%%%%%%%%%%%%%%%%%%%%%%%%%%%%%%%%%%%%%%%%%%%%%%
Within the 2HDM, the search for charged Higgs bosons has mainly proceeded in two channels, $H^+\to\tau^+\nu$ and $H^+\to t\bar b$. In the $\tau\nu$ channel, the searches have already excluded low values of $\tan\beta$ for Yukawa Models~I and X and low masses, $M_{H^\pm}<m_t$. At higher masses, this channel is not yet competitive.

In the intermediate-mass region, $m_t<M_{H^\pm}<480~\text{GeV}$, searches in the $t\bar b$ channel exclude the 2HDM at low values of $\tan\beta$, $\tan\beta\lsim{\cal O}(3)$. In the high-mass region, $480~\text{GeV}<M_{H^\pm}$, $\tan\beta$ values of order unity are also on the verge of being excluded up to about 600~GeV.

The bosonic decay modes, $H^+\to W^+A$,  $H^+\to W^+H$ and  $H^+\to W^+h$ should also be kept in mind, though the former suffer from limited phase space, and the latter from the vanishing of the coupling in the alignment limit.

%%%%%%%%%%%%%%%%%%%%%%%%%%%%%%%%%%%%%%%%%%%%%%%%%
\section*{References}
%%%%%%%%%%%%%%%%%%%%%%%%%%%%%%%%%%%%%%%%%%%%%%%%%
%%%%%%%%%%%%%%%%%%%%%%%%%%%%%%%%%%%%%%%%%%%%%%%%%%

\end{document}